\documentstyle[epsfig, 12pt]{article}

\def\dalemb#1#2{{\vbox{\hrule height .#2pt
        \hbox{\vrule width.#2pt height#1pt \kern#1pt
                \vrule width.#2pt}
        \hrule height.#2pt}}}

\let\a=\alpha \let\b=\beta \let\g=\gamma \let\d=\delta \let\e=\epsilon
\let\z=\zeta  \let\th=\theta  
\let\l=\lambda \let\m=\mu \let\n=\nu \let\x=\xi  
\let\s=\sigma \let\t=\tau  \let\f=\phi  
 
\let\w=\omega      \let\G=\Gamma \let\D=\Delta \let\Th=\Theta 
\let\X=\Xi  \let\S=\Sigma  \let\Y=\Psi
 
\let\la=\label  
  
\def\nn{\nonumber} \def\bd{\begin{document}} \def\ed{\end{document}}
\def\ds{\documentstyle} \let\fr=\frac \let\bl=\bigl \let\br=\bigr
\let\Br=\Bigr \let\Bl=\Bigl
\let\bm=\bibitem
\let\na=\nabla
\def\tU{{\widetilde U}}
\let\pa=\partial \let\ov=\overline
\def\ie{{\it i.e.\ }}
\newcommand{\be}{\begin{equation}}
\newcommand{\ee}{\end{equation}}
\def\ba{\begin{array}}
\def\ea{\end{array}}
\def\ft#1#2{{\textstyle{{\scriptstyle #1}\over {\scriptstyle #2}}}}
\def\fft#1#2{{#1 \over #2}}
\def\F#1#2{{ F_{#1}^{(#2)} }}
\def\cF#1#2{{ {\cal F}_{#1}^{(#2)} }}

\def\R{{\bf R}}
\def\sst#1{{\scriptscriptstyle #1}}
\def\oneone{\rlap 1\mkern4mu{\rm l}}
\def\e7{E_{7(+7)}}
\def\td{\tilde}
\def\wtd{\widetilde}
\def\im{{\rm i}}
\def\bog{Bogomol'nyi\ }
\newcommand{\ho}[1]{$\, ^{#1}$}
\newcommand{\hoch}[1]{$\, ^{#1}$}
\newcommand{\bea}{\begin{eqnarray}}
\newcommand{\eea}{\end{eqnarray}}
\newcommand{\ra}{\rightarrow}
\newcommand{\lra}{\longrightarrow}
\newcommand{\Lra}{\Leftrightarrow}
\newcommand{\ap}{\alpha^\prime}
\newcommand{\bp}{\tilde \beta^\prime}
\newcommand{\cB}{{\cal B}}
\newcommand{\cO}{{\cal O}}
\newcommand{\vecx}{\vec{x}}
\newcommand{\vecy}{\vec{y}}
\newcommand{\vecp}{\vec{p}}
\newcommand{\vecq}{\vec{q}}
\newcommand{\tr}{{\rm tr} }
\newcommand{\Tr}{{\rm Tr} }
\newcommand{\NP}{Nucl. Phys. }

\newcommand{\cL}{{\cal L}}
\newcommand{\cA}{{\cal A}}
\newcommand{\cD}{{\cal D}}

\def\sst#1{{\scriptscriptstyle #1}}
\def\0{{\sst{(0)}}}
\def\1{{\sst{(1)}}}
\def\2{{\sst{(2)}}}
\def\3{{\sst{(3)}}}
\def\4{{\sst{(4)}}}
\def\5{{\sst{(5)}}}
\def\6{{\sst{(6)}}}
\def\7{{\sst{(7)}}}
\def\8{{\sst{(8)}}}
\def\ve{\varepsilon}
\def\vf{\varphi}
\def\F{\Phi}
\def\wg{\wedge}

\newcommand{\tamphys}{\it 
}

\newcommand{\auth}{AUTHORS}

\def\thb{\bar{\theta}}
\def\Thb{\bar{\Theta}}
\def\barp{\bar{p}}
\def\barq{\bar{q}}
\def\barc{\bar{c}}
\def\bard{\bar{d}}
\def\e{\epsilon}

\def \bi{\bibitem}
\def \la {\label}

\def \l {\lambda}
\def\foot{\footnote}
\def \tl  {{\tilde \l}}
\def \sql {{\sqrt \l}}
\def \adss {$AdS_5 \times S^5$\ }
\newcommand{\rf}[1]{(\ref{#1})}
\def \ov {\over}

\def\th{\theta}
\def\Th{\Theta}
\def\vth{\vartheta}
\def\btheta{{\bar\theta}}
\def\ttheta{{{\tilde\theta}}}
\def\bttheta{{{\bar\ttheta}}}
\def\vth{\vartheta}

\def\ra{\rightarrow}
\def\N{{\cal N}}
\def\F{{\cal F}}
\def\uM{\underline{M}}
\def\uN{\underline{N}}
\def\uP{\underline{P}}
\def\cc{\circ}
\def\eqv{\equiv}

\def\ni{\noindent}

\def\Ep{E^{{}^{(+)}}}
\def\Em{E^{{}^{(-)}}}

\def\Mp{M^{{}^{(+)}}}
\def\Mm{M^{{}^{(-)}}}

\def \ha{{1\ov 2}}

\def\r{\rho}

\def\Y{{\rm Y}}
\def\X{{\rm X}}
\def\tY{\tilde{\rm Y}}
\def\tX{\tilde{\rm X}}
\def\dY{\dot{\rm Y}}
\def\dX{\dot{\rm X}}

\def \J {\mathcal{J}}
\def \del {\partial}

\def\dF{\dot{F}}
\def\dG{\dot{G}}
\def\df{\dot{f}}
\def \E {{\cal E}}
\def \S {{\cal S}}
\def \J {{\cal J}}

\def\ms{\mathcal{S}}
\def\mj{\mathcal{J}}
\def\soj{\fr{\ms}{\mj}}
\def \R {{\bf R}}
\def \om {\omega}
\def \bE {\bar E}
\def \x {{\cal X}}

\def \bi{\bibitem}
\def \la {\label}

\def \l {\lambda}
\def\foot{\footnote}
\def \tl  {{\tilde \l}}
\def \sql {{\sqrt \l}}
\def \adss {$AdS_5 \times S^5$\ }
\def \ov {\over}

\def \varpi {{\rm w}}

\def\thb{\bar{\theta}}
\def\Thb{\bar{\Theta}}
\def\psib{\bar{\psi}}
\def\barp{\bar{p}}
\def\barq{\bar{q}}
\def\barc{\bar{c}}
\def\bard{\bar{d}}
\def\e{\epsilon}

\def\At{\tilde{A}}
\def\Bt{\tilde{B}}
\def\ola{\overleftarrow}
\def\ora{\overrightarrow}
\def\at{\tilde{\a}}

\def\ps{\rlap{\, /}\;\,p }
\def\ks{\rlap{\, /}\;\,k }

\def\gym{g_{YM}}

\def\adot{\dot{a}}
\def\bdot{\dot{b}}

\begin{document}
\overfullrule=0pt
\parskip=2pt
\parindent=12pt
\headheight=0in \headsep=0in \topmargin=0in
\oddsidemargin=0in

\vspace{ -3cm}
\thispagestyle{empty}

\begin{center}

{\Large\bf Scattering on D3-branes
  }

 \vspace{.5cm} { I.Y. Park  }\\
 \vskip 0.2cm

{\it Department of Chemistry and Physics, University of Arkansas
at Pine Bluff \\
Pine Bluff, AR 71601, USA \\
inyongpark05@gmail.com}


\end{center}

 \vspace{0.1cm}

 \begin{abstract}
\ni In a direct open string approach we analyze scattering of
massless states on a stack of D3-branes. First we construct vertex
operators on the D-branes. The 4+6 splitting for the fermionic part
is made possible by inserting appropriately defined projection
operators. With the vertex operators constructed we compute various
tree amplitudes. The results are then compared with the
corresponding field theory computations of the $\N=4$ SYM with
$\a'$-corrections: agreements are found. We comment on applications
to AdS/CFT.

\end{abstract}
\newpage

\setcounter{equation}{0}
\setcounter{footnote}{0}
\setcounter{section}{0}


\section{Introduction}
At the heart of the AdS/CFT are the description methods of D-branes.
They can be described either as a hypersurface where an open string
can end or as a solitonic solution of the closed string theory. In
the open string theory description one can use the $D=4\;\; \N=4$
SYM theory as a leading order approximation to the full open string
description. In particular there has been efforts to compute the
anomalous dimensions of some SYM operators.

Although simple and useful the SYM theory does not contain the
effects of the massive open string modes since the SYM is a leading
order approximation: it may be worth studying a higher order in the
approximation accommodating the effects of the massive modes. A
first step toward this direction has been taken in
\cite{Park:2007ev} where the $\a'$-corrected SYM was considered
 in the regular field
theory approach. One loop scalar four point amplitudes were computed
and the counter-terms that remove the divergence were examined.
Unlike the abelian case where the effective action can be obtained
in a closed form,
 in the non-abelian case one must consider string
theory four-point, five-point, etc, separately, and deduce the field
theory action from the results. It may be useful for that purpose to
know the possible forms of the field theory counter-terms in
advance, which is one of the motivations of the work
\cite{Park:2007ev}.

 As stated there, the string-based technique and the field theory technique should
be mutually guiding. Here we turn to the string world-sheet physics.
Since D-branes are stringy objects it ought to, in general, take the
full open string theory for their complete description. Therefore
how the massive open string modes figure into AdS/CFT (or matrix
theory conjectures for that matter) is an interesting and important
issue. The possible relevance of the open string in the context of
AdS/CFT was discussed e.g., in \cite{Park:1999xz,
Park:2001bm}.\footnote{ Related discussions may be found in
 \cite{DiVecchia:2005vm,Kawai:2007ek}.}
With the comparison with the field theory in mind we study the
scattering of massless states. Although the body of a string lives
in ten dimensions its end points remain {\it on} the D3-branes
before and after the scattering. (We only consider such scattering.)
For the purpose of analyzing such scattering it is necessary to
construct the vertex operators in a direct open string approach: the
boundary state formulation for example can not be applied. Below we
will construct the vertex operators. They come in two multiplets
which we call the "scalar multiplet" and the "gauge multiplet"
respectively. As the name suggests they should respectively
correspond to the scalar multiplet and the gauge multiplet in the
N=2 field theory language. What makes it possible to separate the
scalar multiple from the gauge multiplet (or vice versa) is
insertion of appropriately defined projection operators in various
places. The momenta of the vertex operators will be such that they
have non-zero components only along the D3 brane directions.
Physically speaking, for the branes whose location is fixed this
choice of momenta seems natural. In fact it also follows at an
explicit computational level as a consequence of ensuring the
closure of the vertex operator algebra under susy transformation.
Once they are constructed various tree amplitudes can be easily
computed following the standard procedure. We verify that the field
theory computations at $\a'^2$-order can be recovered by expanding
the corresponding string computations at the same order.
\\

The organization of the paper is as follows. In the next section we
briefly review the boundary conditions of D3 branes in the
Green-Schwarz formulation. We then construct two sets of vertex
operators, the "scalar multiplet" and the "gauge multiplet". In
section 3 we compute various tree amplitudes using the standard
world-sheet techniques and compare the results with the
corresponding amplitudes obtained by using the ${\N=4}$ SYM with the
$\a'$-corrections . By computing the tree graphs we are setting the
ground for the loop computation, which is more interesting and
important for the reasons that we list in the conclusion. There we
also comment on future directions and applications of our results to
AdS/CFT.

\section{vertex operator construction}

In this section we construct the vertex operators in a direct open
string framework. We start with a brief review of the light-cone
gauge to set the notations. The vertex operators are constructed
based on the closure under susy transformations as in the D9-brane
case. The additional task, compared with the D9 case, is that now
one should carry out the (4+6) splitting. For the bosonic
coordinates the splitting is obvious whereas with the fermionic
coordinates it is subtle. As we will see below the fermionic
splitting is accomplished through insertion of some projection
operators. Throughout we mostly follow the conventions of
\cite{gsw}.

\subsection{review of light-cone gauge}

In the Green-Schwarz formulation, the string action is given by
\bea S &=& -\fr{1}{2\pi}\int d^2 \s \; \left(\;\sqrt{-g} g^{\a\b}
       \Pi_\a{}^M \Pi_\b{}^N \eta_{MN}
                  +2 i \e^{\a\b} \pa_\a X^M
                 ( \thb^1 \G_M \pa_\b \th^1
                        -\thb^2 \G_M \pa_\b \th^2) \right. \nn\\
  & & \left.\hspace{1in}  - 2 \e^{\a\b}
                       (\thb^1 \G^M\pa_\a\th^1)(\thb^2 \G_M
                \pa_\b \th^2) \;\right)
\eea
where $g=|\det g_{\a\b}|$ and $\Pi_\a^M=\pa_\a
X^M-i\bar{\th}^A\G^M\pa_\a\th^A$. The 32x32 $\G$-matrices are such
that $\G^M,M\neq 0$ is real and symmetric and $\G^0$ is real and
antisymmetric.

Consider a D3 brane extended along the $(X^1,X^2,X^3)$-directions.
We locate it at the origin of the transverse dimensions, i.e., $X^m$
at $\s=0,\pi$. The boundary conditions for the bosonic coordinates
are such that we impose the Neumann conditions for the world volume
coordinates, $X^{\m}=0$, and Dirichlet for the transverse ones,
$X^4,...,X^9$:
\be \pa_\t X^{m} = 0,\quad \s=0,\pi \ee
\be \pa_\s X^{\m}  = 0,\quad \s=0,\pi \ee
For the fermionic coordinates it is necessary to impose a
constraint,
 \be \th^2 =\G_{4 \cdots 9} \,\th^1,\quad \s=0,\pi
 \label{constraint1}
 \ee
which in turn implies the usual half supersymmetry breaking
condition. After the standard light-cone gauge fixing procedure
 \bea
 \G^{+}\th^{1,2}=0
 \eea
one has the following action,
 \bea
 S&=&-\fr12\int\; (T\pa_\a X^i \pa^\a X^i-\fr{i}{\pi}\bar{S}^a \r^\a\pa_\a
 S^a) \nn\\
  &=&-\fr{1}{2\pi}\int\; (\pa_\a X^i \pa^\a X^i-{i}\bar{S}^a \r^\a\pa_\a
 S^a)
 \eea
where $S\equiv \sqrt{p^+}\;\th$. The mode expansion of the bosonic
coordinates  is
 \bea
 X^\m(\s,\t)&=&x^\m+l^2p^\m\t+il\sum_{n\neq 0} \fr1n \a^\m_n
 e^{-in\t}\cos{n\s} \nn\\
 X^m(\s,\t)&=&R^m+\fr{1}{\pi}\D X^m\s+l\sum_{n\neq 0} \fr1n \a^m_n
 e^{-in\t}\sin{n\s} \nn\\
 \eea
 where $R^m, \D X^m$ are the parameters that are associated with the
 locations of the branes. We locate the branes at the origin of the
  transverse 6-plane.
 For an open string with both ends on the D3-branes
 the transverse coordinates become simpler
 \bea
 X^m(\s,\t)=l\sum_{n\neq 0} \fr1n \a^m_n
 e^{-in\t}\sin{n\s}
 \eea
The mode expansion of the fermionic coordinates  is
 \bea
 S^{1a}&=&\sum_{-\infty}^{\infty}s_n^ae^{-in(\t-\s)}\nn\\
 S^{2a}&=&\sum_{-\infty}^{\infty}\g_{4..9}s_n^ae^{-in(\t+\s)}
 \label{me}
 \eea
 where $\g_{4..9}$ is a 8x8 matrix. Note that $\g^2,...,\g^8$
 are real and antisymmetric matrices while $\g^9$ is an
 identity matrix.
This mode expansion yields
 \bea
 \{S^{Aa}(\s),S^{Bb}(\s')\}=2\pi\d^{ab}\d^{AB}\d(\s-\s')
 \eea

\subsection{supercharges}
 Since we will construct the vertex operators mostly based on their susy
 transformations we first obtain the expressions for the supercharges.
 Care is needed with the boundary conditions/terms. There are two sets of susy transformations.
The first set is
 \bea
 \d\th^1=\fr{1}{\sqrt{2}}\eta^a\quad,\quad
 \d\th^2=\fr{1}{\sqrt{2}}\eta^a \quad,\quad\d X^i=0
 \eea
which yields
 \bea
 Q^a=\sqrt{2p^+}\;s_0^a \quad,\quad
 \eea
 It has the same form as the D9-brane case. The second set of the susy
 transformation has the same form as the D9-brane case as well, but it
 is in terms of modified susy parameters, $\ve$:
 \bea
 \d S=
           -\fr{1}{\sqrt{2p^+}}\r\cdot \pa X^i \g^i\ve  \quad,\quad
 \d X^i =-\fr{i}{\sqrt{2p^+}}\bar{\ve}\g^i S
 \eea
To determine $\ve$, we examine the boundary terms that result from
taking the variation on the bosonic term,
 \bea
 \sim \pa_\s X^i(\e_2\g^iS_1-\e_1\g^i S_2)
 \eea
 The $(i=u)$-terms drop due to Neumann boundary condition.
 Substitution of (\ref{constraint1}) into the above equation leads to
 the susy parameters, $
          \ve=\left(
           \begin{array}{c}
        \e          \\
        -\g_{4,...,9}\e        \\
   \end{array}
          \right)$.
The supercharges for this transformation are
 \bea
 Q^{\dot{a}}
 &=&{\fr{1}{\sqrt{p^+}}}
 \left(\sum_{n}\a_n^i[(\g^i)^T\g_{4..9}]s_{-n}^a\right)
 \eea
where $\a_0^i=p^i=(p^u,0)$. It is a column vector. The supercharges
$Q^a,Q^{\adot}$ satisfies the following algebra:
 \bea
 \{Q^a, Q^b\}&=& 2p^+\d^{ab}\nn\\
 \{Q^a, Q^{\bdot}\}&=& -\sqrt{2}p^u\g^u \g_{4..9}\nn\\
 \{Q^{\adot}, Q^{\bdot}\}&=& 2H\d^{\adot\bdot}
 \eea
 where $H=\fr{1}{2p^+}\left(p^up^u+2\left[\sum_{n=1}^{\infty}
 \a_{-m}^ia_{m}^i+m s_{-m}^a s_m^a\right]\right)$.
Since $\g_{4..9}$ appears frequently it is convenient to define
 \bea
 \g\equiv \g_{4..9}=\g_{4..8}
 \eea
where the second equality holds since $\g^9$ is an 8x8 identity
matrix. $\g$ satisfies
 \bea
 \g^T=-\g\quad,\quad [\g,\g^m]=0=\{\g,\g^u\}\quad,\quad \g^2=-1
 \eea

\subsection{vertex operator}

With the supercharges available we are ready to construct the vertex
operators by requiring closure under susy.\footnote{The Lorentz
transformation can be utilized as well \cite{gsw}. The discussion of
Lorentz invariance in the current case goes parallel to the D9 case.
In particular one can show that $[J^{i-},J^{j-}]=0$ requires the
theory in the critical dimension.} We do that in $k^+=0$ frame as in
the D9 brane case. It turns out that they come in two pairs: we call
them a vector multiplet and a scalar multiplet. With the various
gauges and constraints that we have imposed they should correspond
the $\N=2$ field theory scalar multiplet and the gauge multiplet.
Each pair satisfies the modified susy transformation relations given
in (\ref{vvoa}) and (\ref{svoa}) below, which are analogous to the
corresponding D9-brane relations
 \bea
  &&[\eta^a  Q^a, V_{F}(u,k)]  \approx  V_{B}(\tilde{\z},k)\quad,\quad
  [\eta^a  Q^a,V_{B}(\z,k)]   \approx  V_{F}(\tilde{u},k) \nn\\
  &&[\e^{\adot} Q^{\adot}, V_{F}(u,k)] \approx
 V_{B}(\tilde{\tilde{\z}},k) \quad,\quad
 [\e^{\adot} Q^{\adot}, V_{B}(\z,k)]  \approx
  V_{F}(\tilde{\tilde{u}},k)
 \eea
The wave function $u$ satisfies
 \bea
 &&k^+u^a+k^i\g^i_{a\adot}u^{\adot}=0\quad,\quad
 k^-u^{\adot}+k^i\g^i_{\adot a}u^{a}=0
 \eea
The $\approx$ means that the equalities are up to total
$\t$-derivative terms. The closure of each multiplet is made
possible by inserting the following projection operators,
 \bea
 E_+=\fr12(1+i\g)\quad,\quad E_-=\fr12(1-i\g)
 \eea
In particular they appear in the fermionic parts of the vertex
operators bringing the (4+6) splitting. As a natural trial we choose
momenta such that they have non-zero components only along the D3
brane directions. From a physical standpoint this choice seems
inevitable for the branes whose location is fixed. In fact we will
see that it follows as a consequence of the vertex operator algebra
under susy generators. Let's use the convention that $\m,\n$ are the
brane direction with $u,v=2,3$ and $m,n$ are the transverse
directions. With $k^m=0$ the transverse polarization condition
becomes $k^u\z^u=0$.\\

\ni Defining $k^i=(k^u,0),\; \z^i=(\z^u,0)$ the vector multiplet is
 \bea
 V_{Bg}(\z,k)&=&
 (\z^uB_g^u-\z^-B_g^+)e^{ik\cdot X}\nn\\
 V_{Fg}(u,k)&=&(u^aE_-F_g^a+u^{\dot{a}}
  E_+F_g^{\dot{a}})e^{ik\cdot X}
 \eea
where
 \bea
 B_g^+&=&p^+ \nn\\
 B_g^u&=& (\dot{X}^u-R_g^{uj}k^j)\nn\\
 F_g^{\dot{a}}&=&\fr{1}{\sqrt{2p^+}}[((\g^u)^T \dot{X}^uS_1)^{\dot{a}}
      -((\g^m)^T {{X^m}'}S_1)^{\dot{a}}
      +\fr13 :((\g^i)^TS_1)^{\dot{a}}R_g^{ij}:k^j]\nn\\
 F_g^a&=&  \sqrt{\fr{{p^+}}{2}}\;S_1^a
 \eea
where $R_g^{ij}=\fr14 S_1\g^{ij}S_1$. They satisfy the modified
vertex operator algebra,
 \bea
  &&[\eta^a E_+ Q^a, V_{Fg}(u,k)]  \approx  V_{Bg}(\tilde{\z},k)\quad,\quad
  [\eta^a  E_+ Q^a,V_{Bg}(\z,k)]   \approx  V_{Fg}(\tilde{u},k) \nn\\
  &&[\e^{\adot} E_-Q^{\adot}, V_{Fg}(u,k)] \approx
 V_{Bg}(\tilde{\tilde{\z}},k) \quad,\quad
 [\e^{\adot}E_- Q^{\adot}, V_{Bg}(\z,k)]  \approx
  V_{Fg}(\tilde{\tilde{u}},k) \label{vvoa}
 \eea
The wave function $u$ satisfies
 \bea
 &&k^+u^a+k^u\g^u_{a\adot}u^{\adot}=0\quad,\quad
 k^-u^{\adot}+k^u\g^u_{\adot a}u^{a}=0
 \eea
How the projection operators bring the closure can be seen e.g., in
the computation of
 \be
 [\eta^a E_+ Q^a, V_{Fg}(u,k)]  \approx
V_{Bg}(\tilde{\z},k).
 \ee
One of the commutators yields
 \bea
 [\eta E_+ \sqrt{2p^+}s_0, (u E_+)^{^{\adot}}F^{\adot}]
 &=&\eta E_+\g^u E_-u\; {\dot X}^u+\eta E_+\g^m E_-u\;  {X'}^m \nn\\
 &=& \eta E_+\g^u u \;{\dot X}^u
 \eea
where in the second equality the second term has dropped due to the
presence of the projection operators. Therefore even though there is
${X'}^m$ in $F^{\adot}$, one produces the correct form of $B_g^u$.
\\

\ni For the scalar multiplet, we define $k^i=(k^u,0),\;
\xi^i=(0,\xi^m)$:
 \bea
 V_{Bs}(\xi,k)&=&\xi\cdot B_s e^{ik\cdot X}=
 (\xi^mB_s^m)e^{ik\cdot X}\nn\\
 V_{Fs}(w,k)&=&wF_s e^{ik\cdot X}=(w^aE_-F_s^a+w^{\dot{a}}
 E_+F_s^{\dot{a}})e^{ik\cdot X}
 \eea
where
 \bea
 B_s^m&=& ({X'}^m+R_s^{mj}k^j)\nn\\
 F_s^{\dot{a}}&=&\fr{1}{\sqrt{2p^+}}[((\g^u)^T \dot{X}^uS_1)^{\dot{a}}
 -((\g^m)^T {{X^m}'}S_1)^{\dot{a}}
        -\fr13 :((\g^i)^TS_1)^{\dot{a}}R_s^{ij}:k^j]\nn\\
 F_s^a&=&  -\sqrt{\fr{{p^+}}{2}}\;S_1^a
 \eea
 where $R_s^{ij}=\fr14 S_1\g^{ij}S_1=R_g^{ij}$. Note that compared with the
 fermionic term in $B_g^u$ the corresponding term in $B_s^m$ has an
 opposite sign. (This can be checked by applying on $X'^m$ a Lorentz
 transformation that takes a state whose only non-zero momentum is
 $k^-$ to a state that has $k^+=0$ with other components non-zero.)
 It triggers a few sign differences in the subsequent formulas.
They satisfy the modified vertex operator algebra,
 \bea
  &&[\eta^a E_- Q^a, V_{Fs}(w,k)]  \approx  V_{Bs}(\tilde{\xi},k)\quad,\quad
  [\eta^a  E_- Q^a,V_{Bs}(\xi,k)]   \approx  V_{Fs}(\tilde{w},k) \nn\\
  &&[\e^{\adot}E_+ Q^{\adot}, V_{Fs}(w,k)] \approx
 V_{Bs}(\tilde{\tilde{\xi}},k) \quad,\quad
 [\e^{\adot} E_+Q^{\adot}, V_{Bs}(\xi,k)]  \approx
  V_{Fs}(\tilde{\tilde{w}},k) \label{svoa}
 \eea
The wave function $w$ satisfies
 \bea
 k^+w^a+k^u\g^u_{a\adot}w^{\adot}=0\quad,\quad
 k^-w^{\adot}+k^u\g^u_{\adot a}w^{a}=0
 \eea
One can see that $k^m=0$ is required to ensure for example
 \bea
[\e^{\adot} E_+Q^{\adot}, V_{Fs}(w,k)]  \approx
  V_{Bs}(\tilde{\tilde{\xi}},k)
 \eea

\section{Tree level scattering}
The respective closure of the scalar multiplet and the vector
multiplet is already a strong indication that the construction is
correct. We substantiate the claim by computing various tree
amplitudes with the vertex operators just constructed. For the
vector vertex operator the computations essentially the same as the
corresponding computations in the D9-branes. The results are then
expanded at $\a'^2$-order and compared with the corresponding
computations in the $\N=4$ SYM with the $\a'$-corrections.
Agreements are found between the two computations.

\subsection{string computation}
 Consider the
vector three point tree graph. Only the cyclically inequivalent
permutations are added. The computation is precisely analogous to
the D9-brane case yielding
 \bea
 A(VVV)&=& g\;\tr(\l^a\l^b\l^c) <\z^1,k^1|V_g(\z^2,k^2)|\z^3,k^3>
               +((1,a)\leftrightarrow (3,c)) \nn\\
        &=& g\;\tr(\l^a\l^b\l^c)(\z^1\cdot k^2\;\z^2\cdot \z^3+\z^2\cdot k^3\;\z^3\cdot \z^1
        +\z^3\cdot k^1\;\z^1\cdot \z^2)
        +((1,a)\leftrightarrow (3,c)) \nn\\
        &=& 2g\;\tr(\l^a\l^b\l^c)(\z^1\cdot k^2\;\z^2\cdot \z^3+\z^2\cdot k^3\;\z^3\cdot \z^1
        +\z^3\cdot k^1\;\z^1\cdot \z^2)\nn\\
        &=& 2igN f^{abc}\;(\z^1\cdot k^2\;\z^2\cdot \z^3+\z^2\cdot k^3\;\z^3\cdot \z^1
        +\z^3\cdot k^1\;\z^1\cdot \z^2)
 \eea
 where in the fourth equality we have adopted a normalization
 $\Tr\l^a\l^b=2\d^{ab}$.
There is no three point scalar vertex in the $\N=4$ SYM with
$\a'$-corrections. The string scalar three point graph indeed
produces a vanishing result:
 \bea
 A(\f\f\f)&=& g\;\tr(\l^a\l^b\l^c)<\xi^1,k^1|V_s(\xi^2,k^2)|\xi^3,k^3> \nn\\
        &=& g\;\tr(\l^a\l^b\l^c)<\xi^1,k^1|\xi_2^m(X'm+R^{mv}k_2^v)e^{ik_2\cdot
        X}|\xi^3,k^3>\nn\\
        &=& g\;\tr(\l^a\l^b\l^c)\;\d(k_1+k_2+k_3)<\xi^1|\xi_2^m R^{mv}k_2^v e^{ik_2\cdot
        X}|\xi^3>\nn\\
        &=&g\;\tr(\l^a\l^b\l^c)\;\d(k_1+k_2+k_3)\; \xi_2^m k_2^v (\xi_1^m\xi_3^v-\xi_3^m\xi_1^v)=0
 \eea
In the third equality we have used the fact that $X^{'m}$ does not
have a zero mode. Proceeding as in the vector case one gets the
fourth equality which is zero since $\xi^v=0$ for the scalar state.
Similarly the vector-vector-scalar vertex can be shown to vanish
which is consistent with the field theory. The last example of three
point function that does not invlove an external fermonic state is
the vector-scalar-scalar vertex,
 \bea
 A(V\f\f)&=& g\;\tr(\l^a\l^b\l^c)<\xi^1,k^1|V_g(\z^2,k^2)|\xi^3,k^3>
               +((1,a)\leftrightarrow (3,c))\nn\\
         &=& g\;\tr(\l^a\l^b\l^c)\;\z^2\cdot(k^3-k^1)\xi^1\cdot
         \xi^3
 \eea
Our final example of three point amplitude is $A(\psi\psi A_\m)$,
 \bea
 A(\psi\psi A_\m)&=&g\;\tr(\l^a\l^b\l^c)<u_1,k^1|V_f(u_2,k^2)|\z^3,k^3>
               +((1,a)\leftrightarrow (2,b))\nn\\
                 &=&g\;\tr(\l^a\l^b\l^c){u_1}\g^\m E_-u_2\;\z_3^\m
 \eea
where $\g^\m$ for example is an eight by eight matrix. The index
$\mu$ has appeared as a result of covariantizing the index $v$.\\

\ni We turn to the four point amplitudes. For the four vector
amplitude, one gets
 \bea
 A(VV\D VV)&=& \fr{g^2}{2}\;\tr(\l^a\l^b\l^c\l^d)<\z^1,k^1|V_g(\z^2,k^2)
                    \D V_g(\z^3,k^3)|\z^4,k^4>\nn\\
        &=& \fr{g^2}{2}\;\tr(\l^a\l^b\l^c\l^d)<\z^1|
       (1+t/2)\z^2\cdot \z^3 B(1-s/2,-1-t/2)
               \nn\\
&& +\left[-\z^2\cdot k^1\z^3\cdot
k^4-R_0^{uv}(\z_2^uk_2^v\;\z^3\cdot k^4-\z_3^uk_3^v\;\z^2\cdot
k^1)\right.\nn\\
 &&\left. +R_0^{uv}R_0^{u'v'}\z_2^uk_2^v\z_3^{u'}k_3^{v'}\right]B(-s/2,1-t/2)
 \nn\\
 &&+\left[\z^2\cdot k^3\;\z^3\cdot k^4+\z^2\cdot k^1\;\z^3\cdot k^2
 +\z^2\cdot k^3\;\z^3\cdot k^2 \right.\nn\\
 &&\left.+R_0^{uv}(\z_2^u k_2^v\;\z^3\cdot k^2
  -\z_3^u k_3^v\;\z^2\cdot k^3
  -k_2^u k_3^v\;\z^2\cdot \z^3 \right.\\
 &&\left.-\z_2^u\z_3^v\;k^2\cdot k^3
 +k_2^u \z_3^v\;\z^2\cdot k^3
 +\z_2^u k_3^v\;\z^3\cdot k^2 )\right]B(1-s/2,-t/2)
 |\z^4>\nn
 \eea
where
 \bea
 s=-(k_1+k_2)^2 \quad t=-(k_2+k_3)^2 \quad u=-(k_1+k_3)^2
 \eea
This is valid up to the cyclically inequivalent permutations which
will be added below. Compared with the D9-brane case there are some
sign flips which are due to different conventions from \cite{gsw}.
They do not persist in the final form of the amplitude given below.
Using the following identities
 \bea
 <\z^1|\z^4>&=& \z^1\cdot \z^4 \nn\\
 <\z^1|R_0^{uv}|\z^4>&=&-\z_4^u\z_1^v+\z_1^u\z_4^v \nn\\
 <\z^1|R_0^{uv}R_0^{u'v'}|\z^4>&=&\z_1^v\z_4^{u'}\d^{uv'}
 -\z_1^u\z_4^{u'}\d^{vv'}-\z_1^v\z_4^{v'}\d^{uu'}+\z_1^u\z_4^{v'}\d^{vu'}
 \eea
 one can derive
 \bea
  A(VV\D VV)=-\fr{g^2}{2}\fr{\G(-s/2)\G(-t/2)}{\G(1-s/2-t/2)}K
 \eea
where
 \bea
 K=&&-\fr14(st\;\z_1\cdot \z_3\;\z_2\cdot \z_4+
     su\;\z_2\cdot \z_3\;\z_1\cdot \z_4
     +tu\;\z_1\cdot \z_2\;\z_3\cdot \z_4)\nn\\
 &&+\fr12 s(\z_1\cdot k_4\;\z_3\cdot k_2\; \z_2\cdot \z_4
           +\z_2\cdot k_3\;\z_4\cdot k_1\; \z_1\cdot \z_3\nn\\
   &&\quad\quad        +\z_1\cdot k_3\;\z_4\cdot k_2\; \z_2\cdot \z_3
           +\z_2\cdot k_4\;\z_3\cdot k_1\; \z_1\cdot \z_4)\nn\\
&&+\fr12 t(\z_2\cdot k_1\;\z_4\cdot k_3\; \z_3\cdot \z_1
           +\z_3\cdot k_4\;\z_1\cdot k_2\; \z_2\cdot \z_4\nn\\
   &&\quad\quad        +\z_2\cdot k_4\;\z_1\cdot k_3\; \z_3\cdot
   \z_4
           +\z_3\cdot k_1\;\z_4\cdot k_2\; \z_2\cdot \z_1)\nn\\
&& +\fr12 u(\z_1\cdot k_2\;\z_4\cdot k_3\; \z_3\cdot \z_2
           +\z_3\cdot k_4\;\z_2\cdot k_1\; \z_1\cdot \z_4\nn\\
   &&\quad\quad        +\z_1\cdot k_4\;\z_2\cdot k_3\; \z_3\cdot
   \z_4
           +\z_3\cdot k_2\;\z_4\cdot k_1\; \z_1\cdot \z_2)
 \eea
 It has precisely the same form as the D9-brane case
\cite{gsw}. For a small $\a'$-expansion note that
 \bea
 \fr{\G(-s/2)\G(-t/2)}{\G(1-s/2-t/2)}=\fr{4}{st}-\fr{\pi^2}{6}
 +\cdots
 \eea
 The leading terms
in the small $\a'$-expansion are
 \bea
&& \fr{4}{st}\left[\Tr(\l^a\l^b\l^c\l^d)+\Tr(\l^a\l^d\l^c\l^b)\right]K\nn\\
+&&\fr{4}{ut}\left[\Tr(\l^a\l^c\l^b\l^d)+\Tr(\l^a\l^d\l^b\l^c)\right]K\nn\\
+&&\fr{4}{su}\left[\Tr(\l^a\l^b\l^d\l^c)+\Tr(\l^a\l^c\l^d\l^b)\right]K
 \label{v4ptleading}
 \eea
The next to leading order terms come at $l^4$ order:
 \bea
 &&-l^4\fr{g^2}{2} \left(-\fr{\pi^2}{6}\right)K\;\Tr(\l^a\l^b\l^c\l^d+\mbox{5 more
 terms})\nn\\
 =&& 2\pi^2g^2\a'^2\;\mbox{S}\!\Tr(\l^a\l^b\l^c\l^d) \;K
  \eea
which is the same as the field theory result since
$\l^a=\sqrt{2}\;T^a$ where $T^a$ is a generator that is used in the
field theory lagrangian. As an example that does not have a
counter-part in the D9 case consider the vector-vector-scalar-scalar
amplitude: it turns out to be,
 \bea
 A(\f V\D V\f)&=& \fr{g^2}{2}\;\tr(\l^a\l^b\l^c\l^d)<\xi^1,k^1|V_g(\z^2,k^2)
                  \D  V_g(\z^3,k^3)|\xi^4,k^4> \nn\\
             &=& \fr{g^2}{2}\;\tr(\l^a\l^b\l^c\l^d)<\xi^1|
       (1+t/2)\z^2\cdot \z^3 B(1-s/2,-1-t/2)
               \nn\\
&& +\left[-\z^2\cdot k^1\z^3\cdot
k^4-R_0^{uv}(\z_2^uk_2^v\;\z^3\cdot k^4-\z_3^uk_3^v\;\z^2\cdot
k^1)\right.\nn\\
 &&\left. +R_0^{uv}R_0^{u'v'}\z_2^uk_2^v\z_3^{u'}k_3^{v'}\right]B(-s/2,1-t/2)
 \nn\\
 &&+\left[\z^2\cdot k^3\;\z^3\cdot k^4+\z^2\cdot k^1\;\z^3\cdot k^2
 +\z^2\cdot k^3\;\z^3\cdot k^2 \right.\nn\\
 &&\left.+R_0^{uv}(\z_2^u k_2^v\;\z^3\cdot k^2
  -\z_3^u k_3^v\;\z^2\cdot k^3
  -k_2^u k_3^v\;\z^2\cdot \z^3 \right.\\
 &&\left.-\z_2^u\z_3^v\;k^2\cdot k^3
 +k_2^u \z_3^v\;\z^2\cdot k^3
 +\z_2^u k_3^v\;\z^3\cdot k^2 )\right]B(1-s/2,-t/2)
 |\xi^4>\nn
 \eea
After some algebra one can show that the leading term in the
$\a'$-expansion is given by
  \bea
 Ng^2(f^{eab}f^{ecd}+f^{eac}f^{ebd})\;\z_2\cdot \z_3\;\xi_1\cdot \xi_4
 \eea
The next order term can be computed similarly to the four vector
case. It is simpler due to the fact that $\xi\cdot k=0=\xi\cdot \z$.
One gets
 \bea
 2\pi^2g^2 \a'^2\mbox{S}\!\Tr(\l^a\l^b\l^c\l^d)\xi_1\cdot \xi_4
 \left[-\fr{su}{4}\z_2\cdot \z_3+\fr{s}{2}
 (\z_2\cdot k_4\;\z_3\cdot k_1)+\fr{u}{2}
 (\z_3\cdot k_4\;\z_2\cdot k_1)\right]
 \eea
  Concerning the cyclic symmetry it is not present when
there is a mixture of scalar-vertex operators and vector-vertex
operators: it is inconsistent with the broken Lorentz symmetry of
the D3 brane configuration. Our final example of four point
amplitude is four scalar scattering. One gets
 \bea
 \fr{g^2}{2}\a'^2\;\Tr(\l^a\l^b\l^c\l^d)
\fr{\G(-s/2)\G(-t/2)}{\G(1-s/2-t/2)}
 \left(su\;\xi_1\cdot \xi_4\;\xi_2\cdot \xi_3
 +tu\;\xi_1\cdot \xi_2\;\xi_3\cdot \xi_4
 +st\;\xi_2\cdot \xi_4\;\xi_1\cdot \xi_3
 \right)\nn\\
 \eea
The inequivalent cycling order is to be understood. At $\a'^2$ order
it yields
 \bea
  -\fr12\pi^2\a'^2\gym^6\mbox{Str}(\l^a\l^b\l^c\l^d)
  \left(su\;\xi_1\cdot \xi_4\;\xi_2\cdot \xi_3
 +tu\;\xi_1\cdot \xi_2\;\xi_3\cdot \xi_4
 +st\;\xi_2\cdot \xi_4\;\xi_1\cdot \xi_3
 \right)
 \eea
after taking the cycling into account.

\subsection{field theory computation}
In this section we compute the $\a'$-corrections to various
scattering amplitudes in the regular field theory approach. The
normalization of the field theory amplitude is such that one should
multiply $\fr{N}{\gym^2}$ to compare with the string theory.
 Also there could be difference in factors of $i$ which is due to the Wick
rotation in some string computations and the lack thereof in the
corresponding field theory computations.

For the SYM action we take
 \bea
  {\cal L}_{SYM}&=&\left[
   -\fr14 F^a_{\m\n}F^{a\m\n}
 -\fr12 \left(\pa_\m\phi_i^a+f^{abc}A_\m^b\phi_i^c\right)^2
  -\fr12\bar{\psi}^a\G^{\m}\left(\pa_\m\psi^a+f^{abc}A_\m^b\psi^c\right)
  \right.\\
  && \left. -\fr12\;f^{abc}\;\psib^a\G^i\phi_i^b\;\psi^c
  -\fr14 \sum_{i,j}f^{abc}f^{ade}\phi_i^b\phi_j^c\phi_i^d\phi_j^e
   -\fr{1}{2}(\pa_\m A_a^\m)^2-\fr12\pa_\m\w_a^*\left(\pa^\m\w_a+f^{abc}A_b^\m\w_c\right)\right]
 \nn
 \eea
The comparison of the three point amplitudes that only include
bosons is straightforward. For example the vector three point
amplitude is given by
 \bea
 A(VVV)=2\gym^4 f^{abc}\;(\z^1\cdot k^2\;\z^2\cdot \z^3+\z^2\cdot k^3\;\z^3\cdot \z^1
        +\z^3\cdot k^1\;\z^1\cdot \z^2)
 \eea
The comparison of fermionic amplitudes are less trivial. The reason
is that the conventions of the SYM action above are such that the
fermionic fields have 32-components (while it has four dimensional
space-time dependence). For example the $(\psi\psi
A_\m)$-computation yields
 \bea
 A(\psi\psi A_\m)=\fr{i}{2}\gym^4 f^{abc}\fr{1}{k_3^2}
 \fr{1}{\G^0\ks_1}\G^0\G^\m \fr{1}{\G^0\ks_2}
 +((1,a)\leftrightarrow (2,b))
 \eea
Implementing the reduction procedure one gets
 \bea
  A(\psi\psi A_\m)=\gym^4f^{abc}\;{\cal U}_1\G^0\G^\m {\cal U}_2\;\z_3^\m
 \eea
where ${\cal U}_{1,2}$ are 32-component spinors. Being a
Mayorana-Weyl spinor, they can be reduced to 16-component spinors,
$U_{1,2}$, rendering the above expression
 \bea
  A(\psi\psi A_\m)=-\gym^4f^{abc}\; U_1\G^\m  U_2\;\z_3^\m
 \eea
 where the minus sigh came from the 32x32 gamma matrix, $\G^0.$
As further dimensional reduction one keeps only the lower half
components for $U_1$ and upper half for $U_2$,
 \bea
  A(\psi\psi A_\m)&=&-\gym^4f^{abc}\;  w_1{\g^\m}^T   w_2\;
                      \z_3^\m \nn\\
                  &=&\gym^4f^{abc}\;  w_1{\g^\m}  w_2\;
                      \z_3^\m
 \eea
where $\g^\m$ is an 8x8 matrix. Matching with the string theory
identifies
 \bea
 w_{1,2}=E_- u_{1,2}
 \eea
Four point amplitudes are in order. In the leading order the vector
four point function is
 \bea
 -i\gym^6&&(\z_1\cdot \z_3\;\z_2\cdot \z_4\;f^{eab}f^{ecd}
 -\z_1\cdot \z_4\;\z_2\cdot \z_3\;f^{eab}f^{ecd}\nn\\
&& +\z_1\cdot \z_2\;\z_3\cdot \z_4\;f^{eac}f^{ebd}
 -\z_1\cdot \z_4\;\z_3\cdot \z_2\;f^{eac}f^{ebd}\nn\\
&& +\z_1\cdot \z_3\;\z_2\cdot \z_4\;f^{ead}f^{ecb}
 -\z_1\cdot \z_2\;\z_3\cdot \z_4\;f^{ead}f^{ecb})
 \eea
The next order result is
 \bea
 8\pi^2\a'^2\; i\gym^6 \;K \mbox{S}\!\Tr(T^aT^bT^cT^d)
 \eea
which matches the corresponding string theory computation. In the
leading order $<A_\m^a(x_1)\f_k^b(x_2)\f_k^c(x_3)A_\s^d(x_4)>$
yields
 \bea
i\gym^6(f^{eab}f^{ecd}+f^{eac}f^{ebd})\;\z_2\cdot \z_3\;\xi_1\cdot
\xi_4
 \eea
To compare with the string results of the previous section we need
the $\a'$-corrections to the SYM. They were obtained in ten
dimensions
\cite{Gates:1986is,Bergshoeff:1986jm,Tseytlin:1999dj,Cederwall:2001bt,
Bergshoeff:2001dc,Koerber:2001uu}. We keep the $\a'^2$-order terms
and reduce it to four dimensions. The complete list of the terms at
$\a'^2$-order were presented in \cite{Park:2007ev}. Here we quote
only the terms that are relevant for the present computations. For
the four vector scattering it is essentially the same as the D9 case
so we will not repeat here. The vertices for the two scalar and two
vector scattering are
 \bea
(2\pi\a')^2\;\;\mbox{Str}\left[
-\fr{1}{8}F_{\m\n}F^{\m\n}D_\r\phi_kD^\r\phi^k
 -\fr{1}{2} D_\n\phi_iF_{\n\r}F^{\r\s}D^\s\phi^i \right]
 \eea
It is straightforward to show that they yield
 \bea
 4i\pi^2g^2 \a'^2\gym^6\mbox{S}\!\Tr(T^aT^bT^cT^d)\xi_1\cdot \xi_4
 \left[-\fr{su}{2}\z_2\cdot \z_3+{s}
 (\z_2\cdot k_4\;\z_3\cdot k_1)+{u}
 (\z_3\cdot k_4\;\z_2\cdot k_1)\right]
 \eea
which is consistent with the string theory computation. At
$\a'^2$-order the relevant vertices for the four scalar amplitude
are
 \bea
(2\pi\a')^2\;\;\mbox{Str}\left[-\fr{1}{8} D_\m\phi_j D^\m\phi^j
D_\n\phi_kD^\n\phi^k+\fr{1}{4}D_\n\phi_iD^\n\phi^kD_\s\phi_kD^\s\phi^i
 \right]
 \eea
which yields
 \bea
  -2i\pi^2\a'^2\gym^6\mbox{Str}(T^aT^bT^cT^d)
  \left(su\;\xi_1\cdot \xi_4\;\xi_2\cdot \xi_3
 +tu\;\xi_1\cdot \xi_2\;\xi_3\cdot \xi_4
 +st\;\xi_2\cdot \xi_4\;\xi_1\cdot \xi_3
 \right)
 \eea
It again agrees with the previous string computation at the same
order.

\section{Conclusion}

In this article we computed several tree amplitudes. One obvious
future direction is one-loop graphs. With the vertex operators
constructed and tested here we are in a good position to tackle the
problem. The one loop analysis will be presented elsewhere
\cite{progress}.

There are several reasons for the importance of one loop amplitudes.
In the loop computation one expects to face divergence.
 One will need to come up with a regularization how to handle the divergence
in the string theory context. The task will be interesting on its
own right. However, what makes it more so is the possibility that
one might encounter a non-trivial geometry arising while handling
the divergence. (This issue is tied with the question whether/how an
open string attached on a D-brane can feel the gravitational effects
that are produced by the brane.) In e.g., \cite{Gonzalez-Rey:1998uh}
an explicit map was obtained between the quantum (and
non-perturbative) effects and the AdS$_5\times$S$^5$ geometry. There
only the pure SYM part was considered. We expect that the massive
modes will have their contribution to the picture. The work of
\cite{Gonzalez-Rey:1998uh} was in a regular field theory context. It
will be very interesting to see how the geometry arise in the
current set-up of the string world sheet analysis. Perhaps could it
arise through a Fishler-Susskind type mechanism?

The one loop should also be useful to study the string corrections
to the anomalous dimensions of the SYM operators. While $\N=4, D=4 $
SYM theory is a super-renormalizable theory the presence of the new
vertices generates divergence. As well known open superstring yields
finite results for various scattering amplitudes. Therefore it is
natural to expect that there should be a procedure to obtain finite
results from the SYM. The divergence would have to be cancelled by
counter-terms. It will be interesting to see how the way that string
theory deals with divergence is related to that of the field theory.
Once the divergence is removed one will be able to compute the
string corrections to the anomalous dimensions.\footnote{For that
matter one may try to compute the anomalous dimensions directly in
the world sheet framework without detouring to the field theory.} We
will report on thses issues in the future.

\newpage


\begin{thebibliography}{20}



\bibitem{Park:2007ev}
  I.~Y.~Park,
  arXiv:0704.2853 [hep-th].


\bibitem{Park:1999xz}
  I.~Y.~Park,
  Phys.\ Lett.\ B {\bf 468}, 213 (1999)
  [arXiv:hep-th/9907142].


\bibitem{Park:2001bm}
  I.~Y.~Park,
  Phys.\ Rev.\ D {\bf 64}, 081901 (2001)
  [arXiv:hep-th/0106078].



\bibitem{DiVecchia:2005vm}
  P.~Di Vecchia, A.~Liccardo, R.~Marotta and F.~Pezzella,
  Int.\ J.\ Mod.\ Phys.\  A {\bf 20}, 4699 (2005)
  [arXiv:hep-th/0503156].




\bibitem{Kawai:2007ek}
  H.~Kawai and T.~Suyama,
  arXiv:0706.1163 [hep-th].


\bibitem{gsw}
  M. B. Green, J. H. Schwarz and E. witten,
  Superstring theory, vol 1,2, Cambridge Univesrsity Press, 1987


\bibitem{progress}
I.~Y.~Park,
  arXiv:0801.0218 [hep-th].










\bibitem{Gates:1986is}
  S.~J.~Gates, Jr. and S.~Vashakidze,
  Nucl.\ Phys.\  B {\bf 291}, 172 (1987).




\bibitem{Bergshoeff:1986jm}
  E.~Bergshoeff, M.~Rakowski and E.~Sezgin,
  Phys.\ Lett.\ B {\bf 185}, 371 (1987).

\bibitem{Tseytlin:1999dj}
  A.~A.~Tseytlin,
  arXiv:hep-th/9908105.



\bibitem{Cederwall:2001bt}
  M.~Cederwall, B.~E.~W.~Nilsson and D.~Tsimpis,
  JHEP {\bf 0106}, 034 (2001)
  [arXiv:hep-th/0102009]
;
  M.~Cederwall, B.~E.~W.~Nilsson and D.~Tsimpis,
  JHEP {\bf 0107}, 042 (2001)
  [arXiv:hep-th/0104236].






\bibitem{Bergshoeff:2001dc}
  E.~A.~Bergshoeff, A.~Bilal, M.~de Roo and A.~Sevrin,
  JHEP {\bf 0107}, 029 (2001)
  [arXiv:hep-th/0105274].






\bibitem{Koerber:2001uu}
  P.~Koerber and A.~Sevrin,
  JHEP {\bf 0110}, 003 (2001)
  [arXiv:hep-th/0108169].























































































\bibitem{Gonzalez-Rey:1998uh}
  F.~Gonzalez-Rey, B.~Kulik, I.~Y.~Park and M.~Rocek,
  Nucl.\ Phys.\ B {\bf 544}, 218 (1999)
  [arXiv:hep-th/9810152].










































































\end{thebibliography}
\end{document}